\documentstyle[twoside,fleqn,epsfig,espcrc2]{article}
\newcommand{\ee}{\end{enumerate}}        
\newcommand{\bi}{\begin{itemize}}
\newcommand{\ei}{\end{itemize}}

\newcommand{\beq}{\protect\begin{equation}}
\newcommand{\eeq}{\protect\end{equation}}        
\newcommand{\bqa}{\begin{eqnarray}}        
\newcommand{\eqa}{\end{eqnarray}}        
\newcommand{\be}{\begin{enumerate}}

\newcommand{\ie}{{\frenchspacing i.\hspace{0.4mm}e.{}}}

\title{Finite size scaling analysis of compact QED}
\author{
  G.~Arnold, Th.~Lippert, K.~Schilling \\
 {\small Department of Physics, University of
    Wuppertal, D-42097 Wuppertal, Germany}\\
  {Th.~Neuhaus}\\
 {\small Department of Physics, University of
    Aachen, D-52056 Aachen, Germany}}
\begin{document}
\begin{abstract}
  We describe results of a high-statistics finite size scaling
  analysis of 4d compact U(1) lattice gauge theory with Wilson action
  at the phase transition point. Using a multicanonical hybrid Monte
  Carlo algorithm we generate data samples with more than 150
  tunneling events between the metastable states of the system, on
  lattice sizes up to $18^4$. We performed a first analysis within the
  Borgs-Kotecky finite size scaling scheme. As a result, we report
  evidence for a first-order phase transition with a plaquette energy
  gap, $G=0.02667(20)$, at a transition coupling,
  $\beta_T=1.011128(11)$.
\end{abstract}
\maketitle
\section{INTRODUCTION}

The determination of the order of phase transitions is of high
importance for lattice field theories.  For it requires higher than
first-order phase transitions to make contact between lattice and
continuum physics.

In compact quantum electrodynamics (QED) with Wilson action the order
of the transition between the confined and Coulomb phases has been
under debate for many years\cite{KLA97}. On the scale of accessible
lattice sizes the correlation length is large, but the system exhibits
a definite two-peak structure.

In a recent high statistics investigation\cite{KLA97} the latent heat
appeared to decrease with the lattice size $L$, with a critical exponent $\nu$
being neither 0.25 (first order) nor $0.5$ (trivially second order),
with significant subasymptotic contributions at the studied values of
$L \leq 12$.  These findings allow for two scenarios:
\begin{enumerate}
\item The observed double peak structure is a finite size effect and
  vanishes in the thermodynamic limit. The signature of a second-order
  phase transition would eventually emerge at some $L_0 \gg 12$.
\item The phase transition is weakly first-order, \ie, the correlation
  length $\xi$ remains finite, yet large in terms of lattice
  extensions accessible today; this would fake, on small lattices, the
  signature of a second-order transition, since the true value of the
  latent heat would be revealed only in the regime $L>\xi$ .
\end{enumerate}

For a class of spin models with strong first-order phase transitions
finite size scaling theory has become amenable to quantitive studies
through the work of Borgs and Kotecky\cite{BORGS1,BORGS2,BORGS3}.
There the finite volume partition function at temperature $\beta$ in
finite volumes with periodic boundary conditions (neglecting
interfacial contributions) has the remarkably simple form \beq
Z=[e^{-Vf_1(\beta)}+e^{-Vf_2(\beta)+\ln(X)}] \; .
\label{eqpartition}
\eeq The functions $f_1(\beta)$ and $f_2(\beta)$ denote bulk free
energy densities in the two coexisting phases 1 and 2.  $X$ denotes
the asymmetry parameter which is nothing but the relative phase weight
in the probability distribution $P(E)$.

In Ref.\cite{JANKE97} the Borgs-Kotecky ansatz has been extended in a
heuristic manner to the 3d 3-state Potts model undergoing a weakly
first-order phase transition; in this instance detailed consistency
checks have been carried out in order to verify the viability of the
Borgs-Kotecky approach.  Motivated by this success we shall apply in
the following this ansatz to compact QED lattice gauge theory.

\section{SIMULATION DETAILS}
We consider 4d pure U(1) gauge theory with standard Wilson action \beq
S = -\beta\sum_{n,\nu>\mu}\cos(\theta_{\mu\nu}(n)), \eeq where $\beta$
denotes the Wilson coupling and $\theta_{\mu\nu}(n)$ the plaquette
angle. We use a cubic lattice of volume $V=L^4$ with periodic boundary
conditions.

We have implemented three different algorithms for generating the U(1)
gauge field configurations: (a) a local Metropolis (MRS), updating
each link separately, (b) a global hybrid Monte Carlo algorithm (HMC)
and (c) a combination of the multicanonical and the hybrid Monte Carlo
algorithm (MHMC). For details we refer to Refs.
\cite{ARNOLD98,ARNOLD99}. Each update of the complete lattice is
followed by 3 reflection steps\cite{BUNK} to reduce correlation of
successive configurations. The number of generated configurations at
each lattice size $L$ is $\ge 2.5 \times 10^6$.  We additionally
measure the number of tunneling events (flips) as control parameter
for the mobility of the algorithms.

Our simulation parameters and the statistics achieved are listed in
Table \ref{tab:mcit}. As the runs differing by algorithm, HMC
parameters or by weight function are independent we do not recombine
them into one multihistogram. Rather we treat them separately.

\begin{table}[htb]
\caption{Number of configurations generated on different lattice
         sizes. In case of HMC generated data the time step $\Delta t$
         and trajectory length $N_{md}$ is specified. The MHMC runs
         with equal parameter settings differ by the choice of weight parameters.} 
\center
\tiny
\begin{tabular}{|r|l|c|l|r|r|r|}
\hline
L  & $\beta$ & algorithm & $\Delta t$& $N_{md}$  & iterations &flips \\
\hline
\hline
 6& 1.001600    & HMC & .120    &10 & 2500000 & 3581\\  
\hline
 8& 1.007370    & HMC & .093    & 2 & 1250000 & 276\\   
  & 1.007370    & HMC & .093   & 4 & 1250000 & 557\\ 
  & 1.007370    & HMC & .093    & 6 & 1250000 &707 \\
  & 1.007370    & HMC & .093    & 8 & 1250000 &856\\    
  & 1.007370    & HMC & .093    & 9 & 1250000 &913\\
  & 1.007370    & HMC & .093    &10 & 1250000 &932\\      
  & 1.007370    & HMC & .093    &11 & 1250000 &935\\      
  & 1.007370    & HMC & .093    &12 & 1250000 &921\\      
  & 1.007370    & HMC & .093    &13 & 1250000 &935\\      
  & 1.007370    & HMC & .093    &14 & 1250000 &859\\      
  & 1.007370    & HMC & .093    &16 & 1250000 &862\\    
  & 1.007370    & HMC & .093    &13 & 1440000 &1265\\   
\hline
10& 1.009300    & HMC & .071    & 9 & 1000000 &294\\  
  & 1.009300    & HMC & .071    &11 & 1000000 &350\\  
  & 1.009300    & HMC & .071    &15 & 1000000 &351\\  
  & 1.009300    & HMC & .071    &17 & 1000000 &344\\
  & 1.009300    & HMC & .071    &19 & 1000000 &340\\
\hline
12& 1.010143    & MRS  &        &   & 3617000 &571 \\ 
  &             & MHMC & .060        &20 & 1303500 &407 \\   
\hline
14& 1.010598    & MRS  &        &   &  3900000&215\\   
  & 1.010568    & MHMC & .050        &24 &  825200 &186\\   
\hline
16& 1.010753    & MRS  &        &   &  3460000&45 \\
& 1.010753      & MHMC & .045        &26 &   595000&73\\
& 1.010753      & MHMC & .045        &26 &   626000&83\\
& 1.010753      & MHMC & .045        &20 &  1044000&189\\
\hline
18& 1.010900    & MHMC & .042        &28 &   632000&23\\   
  & 1.010900    & MHMC & .042        &28 &   905700&68\\   
  & 1.010900    & MHMC & .042        &28&   980000&61\\   
  & 1.010900    & MHMC & .042        &28&   380000&15\\   
\hline
\end{tabular}
\label{tab:mcit}
\end{table}

\section{MEASUREMENTS}
Based on the plaquette operator
\beq
E=\frac{1}{6V}\sum_{n,\nu>\mu}\cos(\theta_{\mu\nu}(n)).
\eeq
we consider the following cumulants:
\begin{eqnarray}
  C_{v}(\beta,L) & =& %
  6\beta^{2}\,V\left(\left<E^{2}\right> - \left<E\right>^{2}\right), \label{cumulant1}\\
  U_{2}(\beta,L) & = &%
  1 - \frac{\left<E^{2}\right>}{\left<E\right>^{2}},\\
  U_{4}(\beta,L) & = &%
  \frac{1}{3}\left(1-\frac{\left<E^{4}\right>}{\left<E^{2}\right>^{2}}\right).
\label{cumulant3}
\end{eqnarray}
\begin{figure}[!htb]
\centerline{\includegraphics[width=7cm]{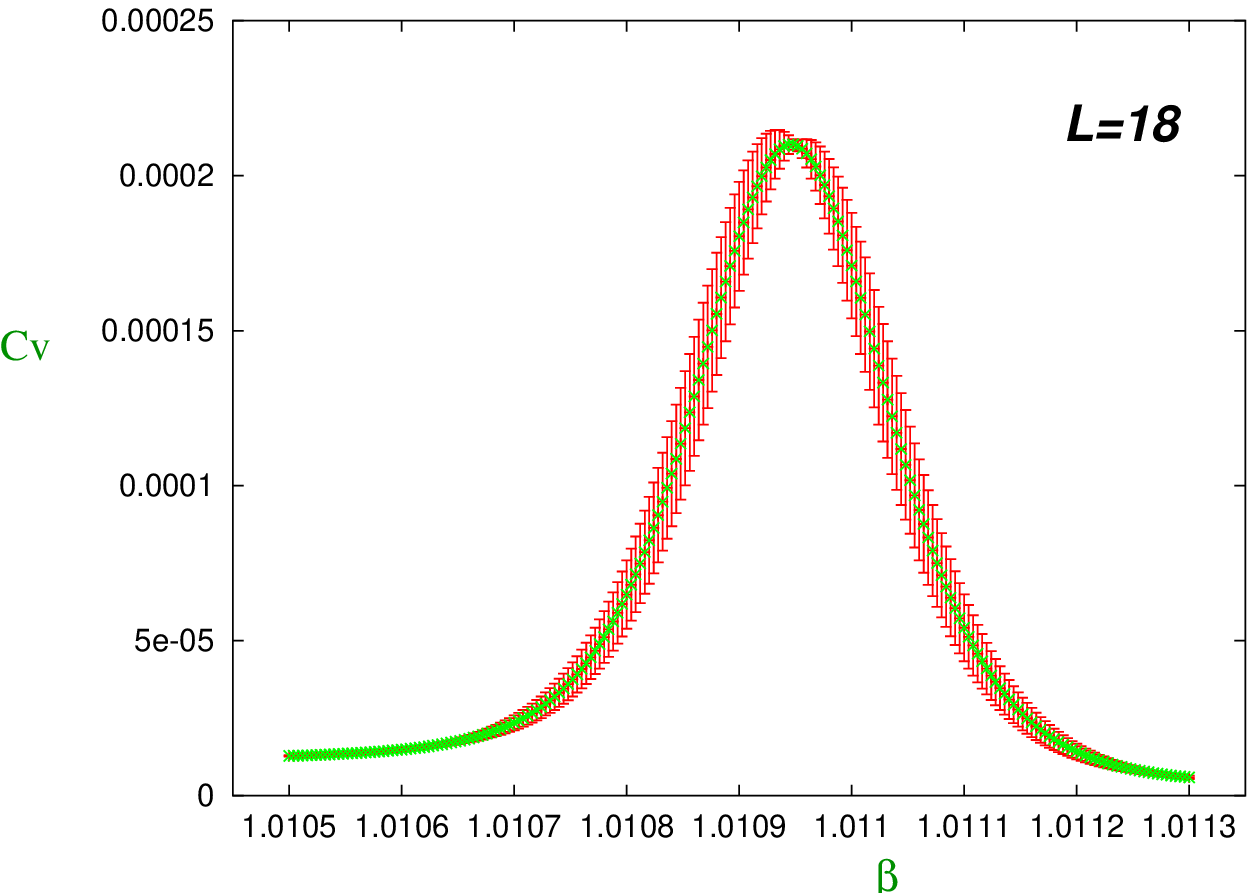}}
\centerline{\includegraphics[width=7cm]{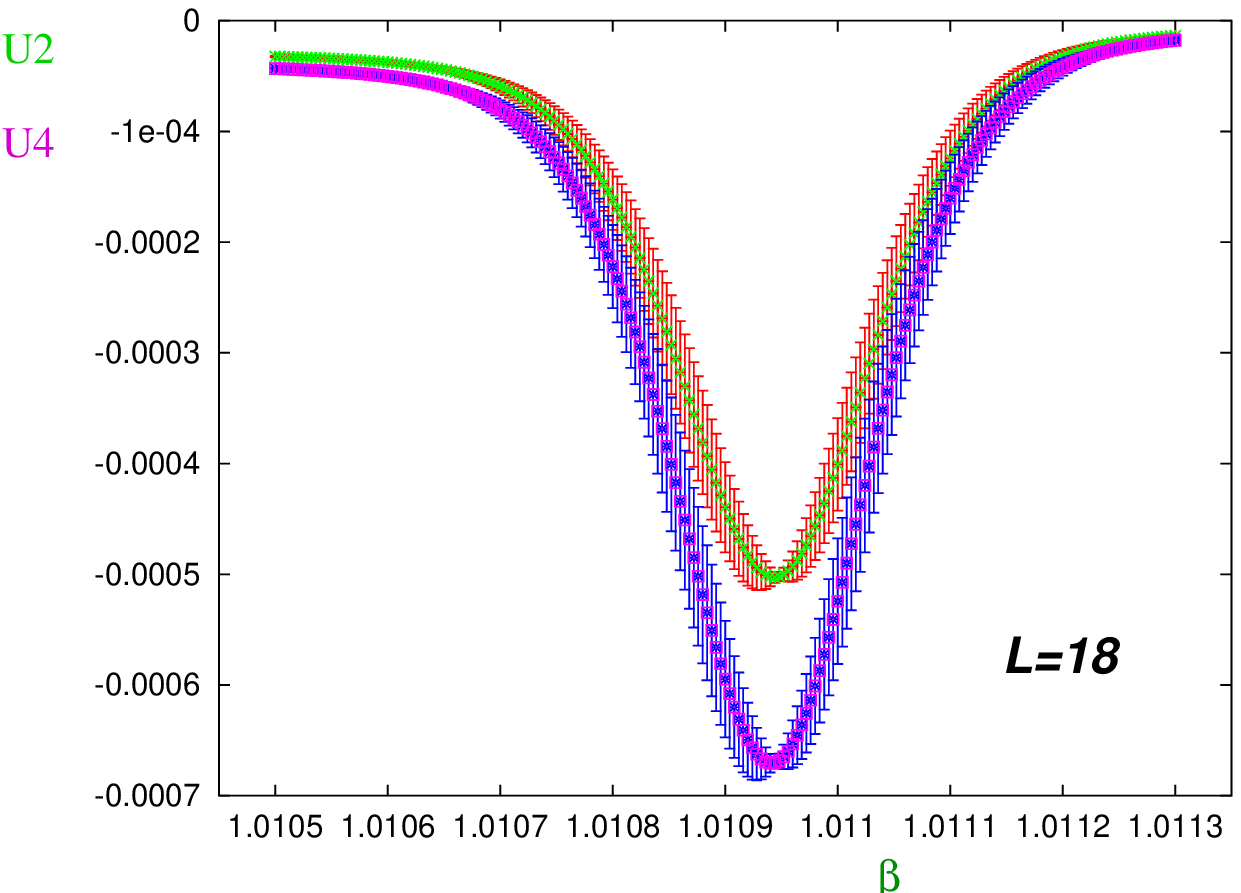}}
\caption{The measured cumulants of our data at $L=18$ taken from the
  third run in table \ref{tab:mcit} reweighted to different couplings
  $\beta$ in order to determine their extrema.}
\label{cumufig}
\end{figure}
The locations $(\beta_c, C_{v,max})$ and $(\beta_c, U_{i,min})$ of
their extrema are determined by reweighting the measured probability
distribution $P(E)$ to different couplings $\beta$ (see Fig
\ref{cumufig}). To calculate the estimates of our cumulants at each
lattice size $L$ we proceed in two steps: i) determine the error of
each individual run performing a jackknife error analysis by
subdivision of the run into ten blocks; ii) calculate the final result
by $\chi^2$-fitting these individual run results to a constant.  The
$L$-dependence of the latter is quoted in Table \ref{tabcmax} and
\ref{tabui}.
\begin{table}[htb]
\caption{Locations of the specific heat maxima as a function of
  lattice size.}   
\center
\small
\begin{tabular}{|r|l|l|}
\hline
L  &  $\beta_c $ & $C_{v,max}/6V \times 10^{-4}$ \\
\hline
\hline
6&1.001794(64)  &9.7728(497)   \\
\hline
8 &1.007413(14)  &5.5428(105)   \\
\hline
10&1.009383(16)  &3.8535(130)   \\
\hline
12&1.010229(14)  &2.9813(118)   \\
\hline
14&1.010626(11)  &2.5470(139)   \\
\hline
16&1.010840(9)   &2.2620(161)   \\
\hline
18&1.010943(8)   &2.1154(152)   \\
\hline
\end{tabular}
\label{tabcmax}
\end{table}

\begin{table}[htb]
\caption{Locations of the $U_i$ minima.}   
\center
\small
\begin{minipage}[t]{7.5cm}
\center
\begin{tabular}{|r|l|l|}
\hline
 L  &  $\beta_{U_2}$ & $U_{2,min}\times 10^{-3}$ \\
\hline
\hline
6&1.001383(63)  &-2.5815(136)\\
\hline
8 &1.007306(14)  &-1.3954(27)\\   
\hline
10&1.009344(16)  &-0.9481(33)\\  
\hline
12&1.010212(14)  &-0.7241(29)\\   
\hline
14&1.010617(11)  &-0.6142(35)\\  
\hline
16&1.010836(9)   &-0.5428(40)\\   
\hline
18&1.010940(8)   &-0.5063(37)\\ 
\hline
\end{tabular}
\begin{tabular}{|r|l|l|}
\hline
L& $\beta_{U_4}$ & $U_{4,min}\times 10^{-3}$\\ \hline
6  &1.001174(63)  &-3.4489(182)\\ \hline
8  &1.007241(14)  &-1.8627(36)\\ \hline
10  &1.009319(16)  &-1.2651(44)\\ \hline
12  &1.010200(14)  &-0.9661(39)\\ \hline
14 &1.010610(11)  &-0.8194(46)\\ \hline
16 &1.010832(9)   &-0.7241(53)\\ \hline
18  &1.010938(8)   &-0.6753(50)\\ 
\hline
\end{tabular}
\end{minipage}
\label{tabui}
\end{table}

\section{FINITE SIZE SCALING ANALYSIS}
Under the assumption that the phase transition is discontinuous, the
Borgs-Kotecky representation of the partition function suggests, that
both, the maxima of the specific heat $C_{v,max}$ and the
pseudocritical $\beta$-values can be expanded in terms of th in the
inverse volume, \beq
\frac{C_{v,max}(V)}{6V}=\frac{1}{4}G^2+\sum_{k=1}^\infty A_kV^{-k}
\label{eqcmax}
\eeq
and
\beq
\beta_c(V)=\beta_T+\sum_{k=1}^\infty B_kV^{-k}. 
\label{eqbetacmax}
\eeq The quantity $G$ stands for the infinite volume gap in the
plaquette energy and $\beta_T$ denotes the infinite volume transition
point of the system. We fit our data for $C_{v,max}$ and $\beta_c$
(given in Table \ref{tabcmax}) to the parameterizations, Eq.
\ref{eqcmax} and \ref{eqbetacmax}.

In order to expose systematic effects in the fit parameter $G$, we
vary the fit range with various ranges $L_{min}\le L \le 18$ as well
as the truncation parameter $k_{max}$.  The results in $G$ are given
in Table \ref{tabfitcmax}. It is possible to apply the
parameterization of Eq. \ref{eqcmax} down to $L_{min}=6$ if one
increases $k_{max}$ such that one degree of freedom remains. It is
nice to see that results with $\chi^2_{dof}<1$ are completely
consistent.

To illustrate the fit stability we additionally included the lowest
expansion coefficients $A_1$ into Table \ref{tabfitcmax}. Note that here again
$\chi^2_{dof}<1$ results exhibit complete stability. 

The observed stability pattern supports the validity of the
$V^{-1}$-expansion. This corroborates the previous analysis
\cite{KLA97} on lattices up to $L=12$.
\begin{table}[htb]
\caption{Fit values of the infinite volume gap $G$ according to
  Eq. \ref{eqcmax}. Fit results in bold face correspond
  to good quality fits ($\chi^2_{dof}<1$).}
\center
\small
\begin{tabular}{|r|l|l|l|l|}
\hline
  $Range$  &   $k_{max}$ &  $\chi^2_{dof}$ &  $G$ &  $A_1$\\
\hline
\hline
14-18 &     1 & 1.03  &    0.02731(17) & 2.62(12)\\
\hline
12-18 &     1 & 6.65  &    0.02779(27) & 2.20(11)\\
   &      2 &\bf 0.50  &\bf  0.02698(18) & \bf 3.22(20)\\
\hline
10-18 &     1 & 25.9  &    0.02837(37) & 1.88(9)\\
   &     2 & 7.76  &    0.02746(22) & 2.61(15)\\
   &     3 &\bf 0.72  &\bf  0.02686(26) &\bf 3.47(34)\\
\hline
8-18  &     1 & 202   &    0.02965(73) & 1.40(8)\\
   &     2 & 7.93  &    0.02788(24) & 2.26(9)\\
   &     3 & 2.21  &    0.02737(22) & 2.72(16)\\
   &     4 &\bf 0.80  &\bf  0.02682(29) &\bf 3.57(41)\\
\hline
6-18  &     1 & 681   &    0.03066(12) & 1.22(12)\\
   &     2 & 113   &    0.02913(60) & 1.64(10)\\
   &     3 & 5.54  &    0.02777(21) & 2.36(9) \\
   &     4 & 2.17  &    0.02735(22) & 2.75(17)\\
   &     5 &\bf 0.83  & \bf    0.02681(31) &\bf 3.60(44)\\
\hline
\end{tabular}
\label{tabfitcmax}
\end{table}
\begin{table}[htb]
\caption{Infinite volume transition coupling $\beta_T$ obtained by
  fitting Eq. \ref{eqbetacmax}. Fit results in bold
  face correspond to good quality fits.}
\center
\small
\begin{tabular}{|r|l|l|l|l|}
\hline
  $Range$  &   $k_{max}$ &  $\chi^2_{dof}$ &  $\beta_T$ & $B_1$\\
\hline
\hline
14-18 &    1&\bf  0.47&\bf  1.0111220(95) &\bf -18.81(56)\\
\hline
12-18 &    1&\bf  0.36&\bf  1.0111169(56) &\bf -18.46(24)\\
   &    2&\bf  0.59&\bf  1.0111238(166)&\bf -19.14(151)\\
\hline
10-18 &     1& 3.62& 1.0110978(124)& -17.39(33)\\
   &     2&\bf 0.83&\bf  1.0111288(123)&\bf -19.65(44)\\
   &     3&\bf 0.67&\bf  1.0111207(196)&\bf -18.71(210)\\
\hline
8-18  &     1&45.5& 1.0110437(370)& -15.13(43)\\
   &     2&\bf 0.44&\bf  1.0111212(80) &\bf -19.08(20)\\
   &     3&\bf 0.36&\bf  1.0111299(82) &\bf -19.79(57)\\
\hline
6-18  &     1& 224 & 1.0109952(785)& -13.99(78)\\
   &     2& 17.8& 1.0110760(244)& -16.72(42)\\
   &     3&\bf 0.29&\bf  1.0111257(45) &\bf -19.44(18)\\
   &     4&\bf 0.36&\bf  1.0111264(85) &\bf -19.81(62)\\
\hline
\end{tabular}
\label{tabfitbetacmax}
\end{table}

As our final result for the infinite volume gap we obtain
\beq
G=0.02698(18)(17).
\label{G1}
\eeq We proceed similarly in fitting the pseudocritical coupling
$\beta_T$, Eq. \ref{eqbetacmax}, with results given in Table
\ref{tabfitbetacmax}. From our analysis of $C_v$ we obtain the
critical Wilson coupling $\beta_{T}=1.011122(10)(8)$.

As a further consistency check we analyze in addition the cumulants
$U_2$ and $U_4$ with parameterizations analogous to Eq.
\ref{eqbetacmax}.
\begin{table}[htb]
\caption{Results for the transition coupling $\beta_T$ obtained by
 analysis of three different cumulants.}
\center
\small
\begin{tabular}{|l|l|}
\hline
 cumulant &  $\beta_T$ \\
\hline
\hline
$C_{v}$ & 1.011122(10)(8) \\
\hline
$U_2$   & 1.011129(14)(8) \\
\hline
$U_4$  & 1.011132(6)(10) \\
\hline
\end{tabular}
\label{tabbetat}
\end{table}
All three estimates (see Table \ref{tabbetat}) are in complete
agreement within their statistical errors. We view this as a
confirmation of the validity of Borgs-Kotecky FSS as applied to
compact QED. As final we quote the average of the values given in
Table \ref{tabbetat}.  \beq \beta_{T}=1.011128(11).
\label{betafinal}
\eeq

It remains to be seen whether our data
exclude the possibility of a continuous singular behaviour with a
vanishing infinite volume gap. Under this alternative assumption one
expects the asymptotic scaling laws
\beq
C_{v,max}(L)=C_1 L^{\frac{2}{\nu}-8}
\label{eqcv2}
\eeq
and 
\beq
\beta_c(L)=\beta_T+a L^{-\frac{1}{\nu}}.
\label{eqbetacv2}
\eeq 
Here $\nu$ denotes the critical exponent of the correlation length  
as hyperscaling is assumed. 

Our data clearly disfavour the validity of Eq. \ref{eqcv2} for any
range of $L$-values. For Eq. \ref{eqbetacv2} however we find the fit
parameters as quoted in Table \ref{tabbeta2}.
\begin{table}[htb]
\caption{Fit results obtained from second-order scaling in
 Eq. \ref{eqbetacv2}.}
\center
\small
\begin{tabular}{|r|l|l|l|}
\hline
 $L_{min}$& $\chi^2_{dof}$ & $\beta_T$ &$\nu^{-1}$ \\
\hline
\hline
6    &   5.12 &   1.011247(26) & 3.20(5)\\
\hline
8    &   2.17 &   1.011204(22) & 3.34(7)\\
\hline
10   &   0.69 &   1.011158(20) & 3.61(11)\\
\hline
12   &   0.64 &   1.011127(32) & 3.90(29)\\
\hline
\end{tabular}
\label{tabbeta2}
\end{table}

Note that fit stability cannot be reached under variation of the fit
intervals $L_{min}\le L \le 18$, neither in $\beta_T$ nor in $\nu$.
However it cannot be excluded, that corrections to scaling could lead
to stable fits also in this scenario.

\section{DIRECT APPROACH TO LATENT HEAT}
\begin{figure}[!b]
\centerline{\includegraphics[width=7cm]{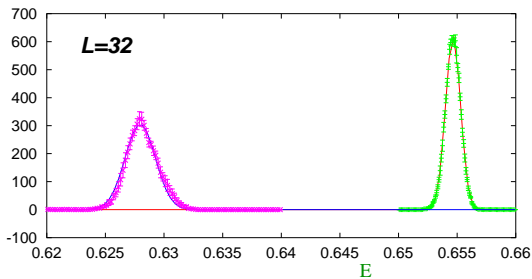}}
\caption{Separate fit of each measured phase peak of the $L=32$ system
  at $\beta_T$ to a Gaussian.}
\label{figgaus}
\end{figure}

\begin{figure}[!tb]
\centerline{\includegraphics[width=7cm]{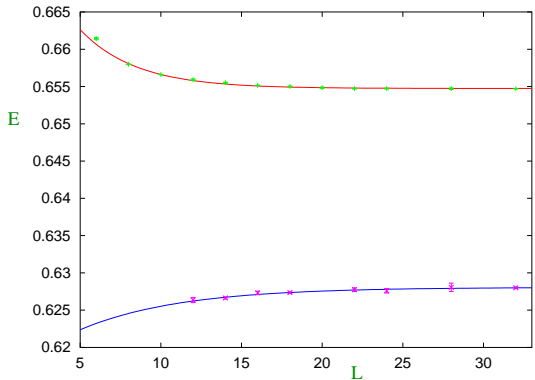}}
\caption{Peak positions $E_i$ as a function of lattice size $L$.}
\label{figcommon}
\end{figure}

\begin{figure}[!htb]
\centerline{\includegraphics[width=7cm]{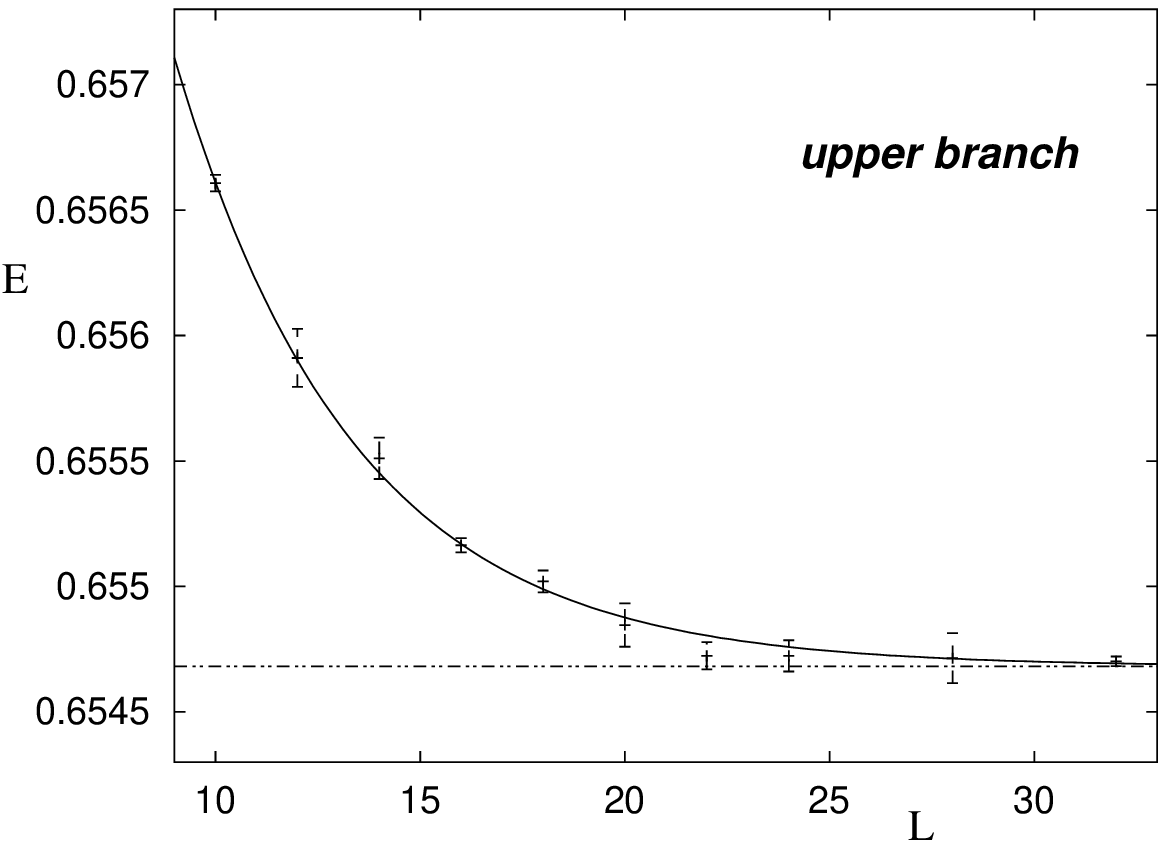}}
\centerline{\includegraphics[width=7cm]{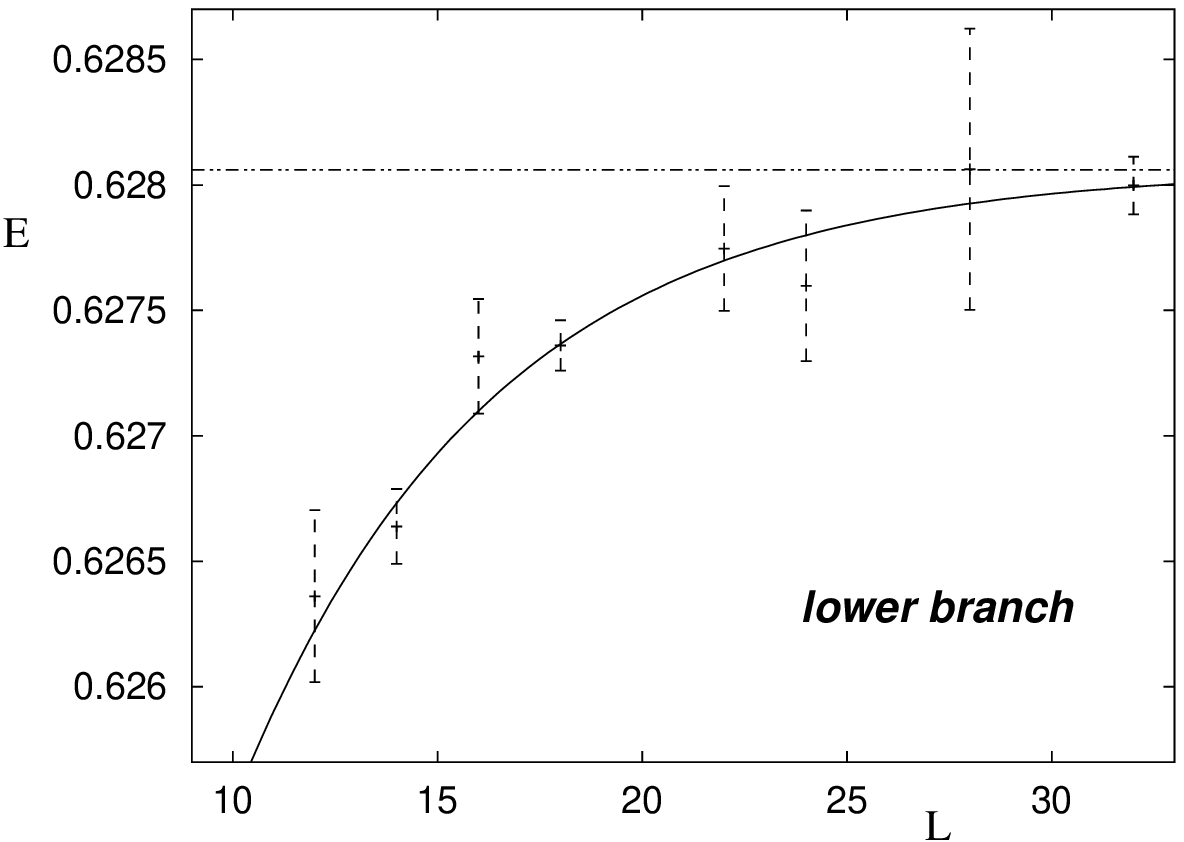}}
\caption{Data fitted to Eq. \ref{branchfit}  on an expanded energy scale.}
\label{figeach}
\end{figure}

Having determined $\beta_T$ very accurately within Borgs-Kotecky FSS
theory we are now in the position to make a 'direct' measurement of
the latent heat based on our determination of $\beta_T$ .  Using the
conventional canonical algorithm we generate $O(10^5)$ configurations
of each metastable phase simulating systems of size
$L=6,8,\ldots,24,28,32$ at $\beta_T$ as given in Eq.  \ref{betafinal}
and locate the positions of the energy peaks in the confined and
Coulomb phase denoted by $E_1$ and $E_2$.  As shown in Fig.
\ref{figgaus} we fit our data for each phase separately to a Gaussian
of the form $a \exp(-b(E-E_i)^2)$. Note that one cannot extract the
relative phase weight from Fig. \ref{figgaus}, because it depicts the
probability distributions $P_i(E)$ of two independent runs $i=1,2$
without any flip.
Fig. \ref{figcommon} shows these peak positions $E_i(L)$ as a function
of lattice size $L$. It is a prediction of the Borgs-Kotecky scheme
that they deviate from their infinite volume value $E_i(\infty)$ by
exponentially small terms only. Accordingly we fit both branches to
\beq E_i(L)=E_i(\infty)+a_i e^{-b_iL},
\label{branchfit}
\eeq achieving high fit quality with $\chi^2_{dof,1}=0.40$ and
$\chi^2_{dof,2}=0.58$. Fig. \ref{figeach} illustrates the data and the
fitted branches on an expanded energy scale.  Due to the wider
distributed peak of the confinement phase the errors of the lower
branch $E_1(L)$ are larger than in case of $E_2(L)$.
With the fit results $E_i(\infty)$ we are in the position to present a
high precision measurement of the infinite volume energy gap, \beq
\hat{G}=E_2(\infty)-E_1(\infty)=0.02667(20).  \eeq

The estimate obtained from first-order scaling of the specific heat as
given in Eq. \ref{G1}  is in agreement with this result. This presents
a further support to the applicability of Borgs-Kotecky FSS to U(1)
gauge theory.

\section{SUMMARY AND CONCLUSIONS}
All cumulants investigated in our high statistics analysis at
$L=6,8,10,12,14,16,18$ are in accord with Borgs Kotecky first-order
FSS.  Our data does not favour second-order FSS as indicated by large
$\chi^2$ and lacking convergence of the critical exponent $\nu$.
Within the framework of Borgs-Kotecky we determined the infinite
volume transition coupling with a relative error of the order of
$10^{-5}$.  A direct investigation of the latent heat at $\beta_T$ for
system sizes up to $L=32$ yields a non-vanishing infinite volume
energy gap $\hat{G}$ with an error of 1\%. The systematic error of the
energy gap due to the error of $\beta_T$ is of the same order as its
statistical error.  The gap $G=0.02698(18)(17)$ extracted from
first-order FSS of the specific heat is in agreement with the direct
measurement of $\hat{G}$.

The material presented here is a progress report. In a forthcoming
paper we shall elaborate further on systematic effects. 

\section*{ACKNOWLEDGMENTS}
This work was done with support from the DFG Graduiertenkolleg
"Feldtheoretische und Numerische Methoden in der Statistischen und
Elementarteilchenphysik".  The computations have been done on a
Connection Machine CM5 at the university of Erlangen, a Cray T3E at
the research center of Juelich and the cluster computer ALiCE at the
university of Wuppertal.

\end{document}